# A Generalized MMSE Detection with Reduced Complexity for Spatially Multiplexed MIMO Signals

Makoto Tanahashi, *Student Member, IEEE*, and Hideki Ochiai, *Member, IEEE*

**Abstract**

In multiple-input multiple-output (MIMO) spatially multiplexing (SM) systems, achievable error rate performance is determined by signal detection strategy. The optimal maximum-likelihood detection (MLD) that exhaustively examines all symbol candidates has exponential complexity and may not be applicable in many practical systems. In this paper, we consider a generalized minimum mean square error (MMSE) detection derived from conditional mean estimation, which in principle behaves equivalently to MLD but also includes a linear MMSE detection as a special case. Motivated by this fact, we propose a low-complexity detection which significantly reduces the number of examined symbol candidates without significant error rate performance degradation from MLD. Our approach is to approximate the probability density function (pdf) of modulated symbols that appears in the exact conditional mean expression such that the decision metric can be cast into a partially closed form. It is found that uniform ring approximation in combination with phase shift keying (PSK) and amplitude phase shift keying (APSK) is promising, as it can achieve a performance even comparable to MLD, while its complexity is linear when the number of transmit antennas is two.

## I. Introduction

Multiple-input multiple-output (MIMO) spatial multiplexing (SM) is a key air-interface technology indispensable for implementing high data-rate wireless communications with limited

The authors are with the Department of Electrical and Computer Engineering, Yokohama National University, 79-5 Tokiwadai, Hodogaya, Yokohama, Kanagawa 240-8501, Japan (e-mail: makoto@ochiailab.dnj.ynu.ac.jp; hideki@ynu.ac.jp)

frequency resources. The enriched capacity of the MIMO-SM systems is simply a consequence of the fact that different signals are transmitted from plural antennas at the same time slot sharing the same frequency band. However, since these transmitted signals travel through the wireless channel where they are linearly superimposed, the receiver must properly unravel the mixture of the signals. Such an operation, variously called demultiplexing, signal detection, or estimation, almost entirely determines achievable performance in terms of error probability, and therefore it has been one of the most important issues in a multi-antenna system design. A number of signal detection frameworks, ranging from simple to sophisticated, have been hitherto proposed [1, 2].

The MIMO signal detection in the simplest form is the linear detection (LD) typically designed based on the zero forcing (ZF) or the minimum mean square error (MMSE) criterion. The LD, though attractive for its simplicity, suffers from limited achievable diversity order which results in poor error rate performance especially when the number of receive antennas is small [1]. Successive interference cancellation (SIC) originated in multi-user code-division multiple access (CDMA) [1] is an extension of LD implementable with incremental complexity, where LD is iteratively performed to replicate and subtract interfering symbols. However, the performance improvement is not significant even at the cost of the delay due to the iterative process. Recently, it has been reported that, when preceded by a nonlinear operation called lattice reduction (LR), LD or SIC can achieve steady signal detection performance even with an ill-conditioned channel matrix [3–5]. The idea behind LR is to transform the lattice from which a signal constellation is drawn, such that the effective channel matrix becomes better-conditioned. The major drawbacks of the LR method include that 1) the complexity of an LR algorithm increases as the channel changes rapidly, and that 2) soft decision is difficult to be made since hard decision is essential on the transformed lattice. The former issue has been addressed with a low-complexity LR algorithm [6] and the latter can be partially overcome by list detection [7], but there remains another limitation that it is not applicable to constellations having no lattice structure such as a phase shift keying (PSK).

While the studies on LD are mainly to assist it for an enhanced performance, those on the exhaustive search techniques such as maximum likelihood detection (MLD) and maximum *a posteriori* (MAP) detection are to reduce their immense search space. For example, in [8, 9] reduced-complexity MLD techniques are presented, where a provisional estimate is first determined with LD and then only its vicinity is exhaustively searched. A similar but more advanced approach is proposed in [10], where the likelihood maximization is done only with addition by exploiting the fact that a single bit flip in a square QAM constellation partially retains the former likelihood. Moreover, QR decomposition (QRD) aided schemes are extensively studied, where the channel matrix is decomposed by QRD so as to create a tree structure on the search space. With this, M-algorithm or sphere decoding (SD) can be employed for efficient search [11], and also several extensions have been proposed to curtail the search space further [12, 13]. It also enables the sequential Monte-Carlo (SMC) technique that can efficiently calculate *a posteriori* probability of symbols [14]. A major drawback common to these QRD-aided schemes is that, due to the sequential structure, multiple symbols cannot be detected in parallel.

In this paper, we focus on a generalized MMSE detection with no linearity constraint, which, to the best of our knowledge, has not been considered in the context of MIMO signal detection. It is literally a nonlinear but optimal processing in the sense that it indeed minimizes mean square error (MSE) by actually computing the exact conditional mean for given received samples. A straightforward implementation of this generalized MMSE detection, however, involves exhaustive summation over all possible symbol patterns as in the case of MLD. Hence, there is no clear advantage over MLD in terms of processing overhead. In spite of this fact, we were inspired by this generalized structure since it has an interesting theoretical aspect from which we develop the idea of a completely new signal detection framework. Specifically, it is known [15] as well as shown in this paper that the generalized MMSE detection reduces to the conventional LD designed with the MMSE principle (i.e., linear MMSE), when all the modulated symbols are hypothetically assumed to be Gaussian distributed. Hence, even though as complicated as MLD in the general form, it also includes the least complicated scheme as a special case. This fact motivated us to derive a new detection scheme that balances performance and complexity, where the key idea is to approximate a selected one (rather than all) of the multiplexed symbols by a continuous random variable (RV) whose distribution should well capture a given shape of the discrete constellation. As a result, the exhaustive summation entailed in the generalized form of the MMSE detection is partially integrated out. In this manner, it is possible to reduce the complexity order from $M^{N_t}$ to $M^{N_t-1}$, where $M$ denotes modulation multiplicity and $N_t$ the number of transmit antennas. The benefit of this complexity reduction becomes apparent when $N_t$ is relatively small: in the case of $N_t = 2$, the complexity reduces to a linear order of $M$, and hence the implementation has now become feasible with limited computational resource.

The rest of this paper is organized as follows: our system model and objective are stated in the next section. In Section III, some existing signal detection methods closely related to our approach are briefly reviewed, followed by the description of our new scheme in Section IV. Section V is devoted to numerical evaluation of error rate performance and Section VI concludes the paper.

We use the following notations: for a vector or a matrix, $^T$ and $^H$ denote its transpose and Hermitian transpose, respectively. For a complex number, $^*$ denotes its conjugate. The entire set of real and complex numbers are denoted by $\mathbb{R}$ and $\mathbb{C}$. An integral in the form of $\int_{\mathbb{C}^N} d^N\boldsymbol{x}$ means an integration over $N$-dimensional complex field, $\boldsymbol{I}_N$ denotes the identity matrix with the dimension of $N$, and $||\boldsymbol{x}|| = \sqrt{\boldsymbol{x}^H\boldsymbol{x}}$ is the norm of a given vector $\boldsymbol{x}$. For RVs which are denoted in upper cases, their lower cases stand for realizations. The functions $P_X(x)$ and $p_X(x)$ represent the probability and pdf of a given RV $X$, respectively. The same notational rule applies to the vector of RVs; i.e., $P_{\boldsymbol{X}}(\boldsymbol{x})$ and $p_{\boldsymbol{X}}(\boldsymbol{x})$ denote the joint probability and joint pdf, respectively, for a vector of RVs $\boldsymbol{X}$. The logarithm for complex numbers always takes the principal branch.

## II. System Model

We consider a MIMO system with $N_t$ transmit and $N_r$ receive antennas ($N_t \leq N_r$) as depicted in Fig. 1. The input-output relationship of the channel is given by

$$\boldsymbol{Y} = \boldsymbol{H}\boldsymbol{X} + \boldsymbol{W}, \tag{1}$$

where

$$\boldsymbol{X} := \begin{bmatrix} X_1 & \cdots & X_{N_t} \end{bmatrix}^T, \tag{2}$$

$$\boldsymbol{Y} := \begin{bmatrix} Y_1 & \cdots & Y_{N_r} \end{bmatrix}^T, \tag{3}$$

$$\boldsymbol{W} := \begin{bmatrix} W_1 & \cdots & W_{N_r} \end{bmatrix}^T, \tag{4}$$

$$\boldsymbol{H} := \begin{bmatrix} \boldsymbol{h}_1 & \cdots & \boldsymbol{h}_{N_t} \end{bmatrix}, \quad \boldsymbol{h}_n := \begin{bmatrix} h_{1,n} & \cdots & h_{N_r,n} \end{bmatrix}^T, \; 1 \leq n \leq N_t \tag{5}$$

are a transmit symbol vector, a receive symbol vector, a noise vector, and a channel matrix, respectively. Assuming no beamforming before transmission, each transmit symbol $X_n$, $n = 1, \ldots, N_t$, is identically independently distributed (i.i.d.) and takes with an equal probability a value from the set of $M$ signal points (i.e, constellation), $\mathcal{X}_M$. Each element in the noise vector is also i.i.d. and follows the complex Gaussian distribution with zero mean and the variance $N_0$.

The task of the signal detector is, for a given event $\boldsymbol{Y} = \boldsymbol{y}$, to calculate the corresponding estimate of transmit vector $\hat{\boldsymbol{X}} := \begin{bmatrix} \hat{X}_1 & \cdots & \hat{X}_{N_t} \end{bmatrix}$, where we assume that $\boldsymbol{H}$, $N_0$, and $\mathcal{X}_M$ are *a priori* knowledge available upon calculating $\hat{\boldsymbol{X}}$ at the receiver.

## III. Signal Detection Methods

### A. Linear Detection (LD)

In LD, its detection operation is restricted to linear arithmetic, and hence $\hat{\boldsymbol{X}}$ can be written in the form of

$$\hat{\boldsymbol{X}} = \boldsymbol{G}\boldsymbol{y}, \tag{6}$$

where the matrix $\boldsymbol{G}$ is adapted to $\boldsymbol{H}$ usually based on the ZF or MMSE principle. In the ZF principle, $\boldsymbol{G}$ corresponds to the pseudo-inverse of $\boldsymbol{H}$:

$$\boldsymbol{G}_{\text{ZF}} = (\boldsymbol{H}^H\boldsymbol{H})^{-1}\boldsymbol{H}^H, \tag{7}$$

which thus completely inverts the linear transformation made in the channel. However, since it adversely amplifies the noise term $\boldsymbol{W}$ at the same time, the following MMSE option is preferable in most cases.

The detection matrix $\boldsymbol{G}$ that satisfies MMSE is by definition written as

$$\boldsymbol{G}_{\text{MMSE}} = \arg\min_{\boldsymbol{G}} E_{\boldsymbol{X},\boldsymbol{W}}\left[||\boldsymbol{X} - \hat{\boldsymbol{X}}||^2\right]. \tag{8}$$

The use of the orthogonality principle is the simplest way to derive $\boldsymbol{G}_{\text{MMSE}}$ satisfying the definition above:

$$E_{\boldsymbol{X},\boldsymbol{W}}\left[(\boldsymbol{X} - \hat{\boldsymbol{X}})\boldsymbol{Y}^H\right] = \boldsymbol{0}, \tag{9}$$

from which $\boldsymbol{G}_{\text{MMSE}}$ can be developed as

$$\boldsymbol{G}_{\text{MMSE}} = E_{\boldsymbol{X},\boldsymbol{W}}\left[\boldsymbol{X}\boldsymbol{Y}^H\right]\left(E_{\boldsymbol{X},\boldsymbol{W}}\left[\boldsymbol{Y}\boldsymbol{Y}^H\right]\right)^{-1} = \boldsymbol{H}^H\left(\boldsymbol{H}\boldsymbol{H}^H + N_0\boldsymbol{I}_{N_r}\right)^{-1}. \tag{10}$$

As shown in Appendix I, this has another form similar to (7) as

$$\boldsymbol{G}_{\text{MMSE}} = \left(\boldsymbol{H}^H\boldsymbol{H} + N_0\boldsymbol{I}_{N_t}\right)^{-1}\boldsymbol{H}^H. \tag{11}$$

### B. Maximum Likelihood and Maximum a Posteriori Detections

In MAP detection, $\hat{\boldsymbol{X}}$ is given as the one replica of possible transmit vectors that maximizes *a posteriori* probability $P_{\boldsymbol{X}|\boldsymbol{Y}}(\boldsymbol{x}|\boldsymbol{y})$:

$$\hat{\boldsymbol{X}} = \arg \max_{\boldsymbol{x} \in \mathcal{X}_M^{N_t}} P_{\boldsymbol{X}|\boldsymbol{Y}}(\boldsymbol{x}|\boldsymbol{y}). \tag{12}$$

Taking logarithm of $P_{\boldsymbol{X}|\boldsymbol{Y}}(\boldsymbol{x}|\boldsymbol{y})$ and performing some manipulations on this formula, it follows that

$$\hat{\boldsymbol{X}} = \arg \min_{\boldsymbol{x} \in \mathcal{X}_M^{N_t}} ||\boldsymbol{y} - \boldsymbol{H}\boldsymbol{x}||^2 - N_0 \log P_{\boldsymbol{X}}(\boldsymbol{x}). \tag{13}$$

Without *a priori* knowledge, all possible $\boldsymbol{x}$ should be assumed to be equally probable. In this case, the second term that depends on $\log P_{\boldsymbol{X}}(\boldsymbol{x})$ can be eliminated so that it reduces to the well-known form of MLD:

$$\hat{\boldsymbol{X}} = \arg \min_{\boldsymbol{x} \in \mathcal{X}_M^{N_t}} ||\boldsymbol{y} - \boldsymbol{H}\boldsymbol{x}||^2. \tag{14}$$

### C. Conditional Mean Estimation

In Section III-A, we saw that liner transformation of $\boldsymbol{y}$ by $\boldsymbol{G}_{\text{MMSE}}$ minimizes MSE. However, with no linearity constraint, the linear MMSE is no longer optimal and the conditional mean estimation expressed as

$$\hat{\boldsymbol{X}} = E_{\boldsymbol{X}}\left[\boldsymbol{X}|\boldsymbol{Y} = \boldsymbol{y}\right] \tag{15}$$

is known to minimize MSE [15]. The formula above suggests simply that the optimal estimate should be the mean of the possible transmit vectors whose corresponding receive vector is $\boldsymbol{y}$, and it is evident that MSE is minimized by this principle.

Let us replace the expectation in (15) explicitly by summation and *a posteriori* probability $P_{\boldsymbol{X}|\boldsymbol{Y}}(\boldsymbol{x}|\boldsymbol{y})$:

$$\hat{\boldsymbol{X}} = \sum_{\boldsymbol{x} \in \mathcal{X}_M^{N_t}} \boldsymbol{x} P_{\boldsymbol{X}|\boldsymbol{Y}}(\boldsymbol{x}|\boldsymbol{y}). \tag{16}$$

This conditional mean estimation expression can be interpreted as a soft-output MAP detection that weights all the possible patterns by their *a posteriori* probabilities, rather than finding the most probable one. Thus, if a hard decision is made on $\hat{\boldsymbol{X}}$, the performance is the same as the MAP detection (also the same as MLD because of the i.i.d. assumption of $\boldsymbol{X}$). In order to actually compute (16), the Bayes' theorem is invoked to rewrite the equation in terms of squared distance as

$$\hat{\boldsymbol{X}} = \frac{\sum_{\boldsymbol{x}\in\mathcal{X}_M^{N_t}} \boldsymbol{x} P_{\boldsymbol{X}}(\boldsymbol{x})\lambda(\boldsymbol{x})}{\sum_{\boldsymbol{x}\in\mathcal{X}_M^{N_t}} P_{\boldsymbol{X}}(\boldsymbol{x})\lambda(\boldsymbol{x})}, \tag{17}$$

where

$$\lambda(\boldsymbol{x}) := \exp\left(-\frac{||\boldsymbol{y}-\boldsymbol{H}\boldsymbol{x}||^2}{N_0}\right). \tag{18}$$

An example of this type of MMSE detection is found in [16, 17].

Upon deriving (16) from (15), we have made use of the fact that each symbol $X_n$, $n = 1, \ldots, N_t$ is a discrete RV having finite realizations. Let us temporarily presume that $X_n$ be a continuous RV having a pdf $p_{X_n}(x_n)$, and rewrite (15) as

$$\hat{\boldsymbol{X}} = \int_{\mathbb{C}^{N_t}} \boldsymbol{x} p_{\boldsymbol{X}|\boldsymbol{Y}}(\boldsymbol{x}|\boldsymbol{y}) d^{N_t}\boldsymbol{x} = \frac{\int_{\mathbb{C}^{N_t}} \boldsymbol{x} p_{\boldsymbol{X}}(\boldsymbol{x})\lambda(\boldsymbol{x}) d^{N_t}\boldsymbol{x}}{\int_{\mathbb{C}^{N_t}} p_{\boldsymbol{X}}(\boldsymbol{x})\lambda(\boldsymbol{x}) d^{N_t}\boldsymbol{x}}. \tag{19}$$

From this formula consisting of multivariate integrals, the following interesting fact results:

*Proposition 1:* If one substitutes a Gaussian pdf for $p_{X_n}(x_n) = \frac{1}{\pi} e^{-|x_n|^2}$ (in vector notation, $p_{\boldsymbol{X}}(\boldsymbol{x}) = \frac{1}{\pi^{N_t}} e^{-||\boldsymbol{x}||^2}$), then the multivariate integrals in (19) are completely solvable, and more importantly, the simplest form of the developed result coincides with $\hat{\boldsymbol{X}} = \boldsymbol{G}_{\text{MMSE}}\boldsymbol{y}$.

*Proof:* See Appendix II. ■

Hence, the conditional mean estimation with Gaussian approximation of symbols reduces to the linear MMSE [15]. One reason for significant performance gap between the linear MMSE and MLD stems from the fact that the former implies approximating the symbol distribution by Gaussian. Since the distribution of the transmitted symbols in general considerably deviates from Gaussian, this approximation may cause a significant discrepancy in the metric calculation.

## IV. A New MMSE Receiver Based on Symbol Distribution Approximation

In the previous section we have shown that, if we keep the symbol distribution intact, then the conditional mean estimation performs as optimally as MLD; on the other hand, it is reduced to the linear MMSE if we approximate the pdf by Gaussian. Motivated by this fact, we explore another approximation that does neither significantly lose a property of original constellation shape, nor require any integration. Specifically, we seek a good approximation of pdf with respect to a single variable in the multivariate integrals, so that the estimation $\hat{\boldsymbol{X}}$ is cast into a partially closed-form. As a result, complexity order is reduced from $M^{N_t}$ to $M^{N_t-1}$.

In what follows, we restrict our attention to the case with $N_t = 2$ since in this case the complexity of MIMO detection reduces to linear order, which is most beneficial in practice. Note that the scenario with $N_t = 2$ is considered to be practical in many MIMO systems, e.g., in the uplink from a small mobile terminal to a base station, where the terminal cannot possess more than two antennas due to size limitation.

In the case of $N_t = 2$, $\boldsymbol{X}$ and $\boldsymbol{H}$ are $2 \times 1$ and $N_r \times 2$, respectively:

$$\boldsymbol{X} = \begin{bmatrix} X_1 & X_2 \end{bmatrix}^T, \tag{20}$$

$$\boldsymbol{H} = \begin{bmatrix} \boldsymbol{h}_1 & \boldsymbol{h}_2 \end{bmatrix}, \quad \boldsymbol{h}_n = \begin{bmatrix} h_{1,n} & \cdots & h_{N_r,n} \end{bmatrix}^T, \, n = 1, 2. \tag{21}$$

In order to detect $\hat{X}_1$ and $\hat{X}_2$ separately in parallel as in Fig. 1, we rewrite (17) element-wise as

$$\hat{X}_n = \frac{\sum_{x_1 \in \mathcal{X}_M} \sum_{x_2 \in \mathcal{X}_M} x_n P_{X_1}(x_1) P_{X_2}(x_2) \lambda(x_1, x_2)}{\sum_{x_1 \in \mathcal{X}_M} \sum_{x_2 \in \mathcal{X}_M} P_{X_1}(x_1) P_{X_2}(x_2) \lambda(x_1, x_2)}, \, n = 1, 2, \tag{22}$$

where

$$\lambda(x_1, x_2) := \exp\left( -\frac{\sum_{k=1}^{N_r} |y_k - (h_{k,1} x_1 + h_{k,2} x_2)|^2}{N_0} \right). \tag{23}$$

It is sufficient to derive an estimation formula with respect only to $\hat{X}_1$; that for $\hat{X}_2$ is given by the same form, except that $\boldsymbol{h}_1$ and $\boldsymbol{h}_2$ are interchanged. Specifically, in element-wise notation, the channel coefficients are exchanged as

$$h_{1,1} \leftrightarrow h_{1,2}, \quad h_{2,1} \leftrightarrow h_{2,2}, \quad \cdots, \quad h_{N_r,1} \leftrightarrow h_{N_r,2}. \tag{24}$$

### *A. Approximation of Desired or Interfering Symbols*

Upon calculating (22), we assume that either $X_1$ or $X_2$ is a RV having continuous pdf $f(x) : \mathbb{C} \to \mathbb{R}$, whereas the other is left intact. If this assumption is made for $X_1$, $\hat{X}_1$ of (22) is expressed as

$$\hat{X}_1 = \frac{\sum_{x_2 \in \mathcal{X}_M} \alpha(x_2)}{\sum_{x_2 \in \mathcal{X}_M} \beta(x_2)}, \tag{25}$$

where

$$\alpha(x_2) := \int_{\mathbb{C}} x \lambda(x, x_2) f(x) dx \tag{26}$$

$$\beta(x_2) := \int_{\mathbb{C}} \lambda(x, x_2) f(x) dx. \tag{27}$$

In this example detector, $X_1$ is the *desired* symbol that is to be estimated. We refer to this type of detection in which the pdf approximation is performed to the desired symbol as a Type-I detection.

Similarly, when this approximation is applied to $X_2$, we obtain

$$\hat{X}_1 = \frac{\sum_{x_1 \in \mathcal{X}_M} x_1 \beta'(x_1)}{\sum_{x_1 \in \mathcal{X}_M} \beta'(x_1)}, \tag{28}$$

where $\beta'(x)$ is equal to $\beta(x)$ defined in (27) with the variable interchanges defined in (24). Note that $X_2$ can be seen as an *interfering* symbol to the detector of $X_1$, and we refer to this type of detection that applies a pdf approximation to the interfering symbols as a Type-II detection.

The function $f(x)$ should meet the following two requirements: 1) it should well approximate the original probability distribution of a signal constellation, and 2) it should transform $\alpha(x)$ and $\beta(x)$ into closed-form expressions. In what follows, we examine several forms of $f(x)$ and derive the corresponding Type-I and II detection formulae, which are summarized in Table I. Note that each formula is given in the log domain, i.e., in a form of $\log \hat{X}_1$ rather than $\hat{X}_1$ itself, in order to avoid computational overflow.

## *B. Preliminaries*

We here define several auxiliary variables such that the subsequent mathematical expressions become concise. In Type-I, we will use

$$w := \boldsymbol{y}^H \boldsymbol{h}_2 = \sum_{k=1}^{N_r} y_k^* h_{k,2}, \tag{29}$$

$$u := ||\boldsymbol{h}_1||^2 = \sum_{k=1}^{N_r} |h_{k,1}|^2, \tag{30}$$

$$v := ||\boldsymbol{h}_2||^2 = \sum_{k=1}^{N_r} |h_{k,2}|^2, \tag{31}$$

$$z := (\boldsymbol{y} - \boldsymbol{h}_2 x_2)^H \boldsymbol{h}_1 = \sum_{k=1}^{N_r} (y_k - h_{k,2} x_2)^* h_{k,1}, \tag{32}$$

$$r_z := |z|, \tag{33}$$

$$\phi_z = \arg z. \tag{34}$$

In Type-II, these variables are interpreted as

$$w := \boldsymbol{y}^H \boldsymbol{h}_1 = \sum_{k=1}^{N_r} y_k^* h_{k,1}, \tag{35}$$

$$u := ||\boldsymbol{h}_2||^2 = \sum_{k=1}^{N_r} |h_{k,2}|^2, \tag{36}$$

$$v := ||\boldsymbol{h}_1||^2 = \sum_{k=1}^{N_r} |h_{k,1}|^2, \tag{37}$$

$$z := (\boldsymbol{y} - \boldsymbol{h}_1 x_1)^H \boldsymbol{h}_2 = \sum_{k=1}^{N_r} (y_k - h_{k,1} x_1)^* h_{k,2}, \tag{38}$$

$$r_z := |z|. \tag{39}$$

Note that $\phi_z$ does not appear in the Type-II formulae. In addition, we denote by $C$ a constant that appears only in the derivation stage.

In the detection formulae expressed in the log domain, we make use of special summations denoted by $\sum^{\mathbb{R},\log}$ and $\sum^{\mathbb{C},\log}$. These are Jacobian logarithm summations in the real and complex domains, respectively, in which the ordinary addition is replaced by other operations as follows:

$$\sum^{\mathbb{R},\log} : \quad a + b \to \max(a,b) + \log(1 + e^{-|a-b|}), \quad a, b \in \mathbb{R} \tag{40}$$

$$\sum^{\mathbb{C},\log} : \quad a + b \to \max_{\mathrm{Re}}(a,b) + \log(1 + e^{-(\max_{\mathrm{Re}}(a,b) - \min_{\mathrm{Re}}(a,b))}), \quad a, b \in \mathbb{C}, \tag{41}$$

where $\max_{\mathrm{Re}}(a,b)$ denotes the one whose real part is greater than the other, and $\min_{\mathrm{Re}}(a,b)$ is defined accordingly. The derivations of these summations are provided in Appendix III. In the above expressions, neglecting the second terms with $\log$ and $\exp$ operations results in significant reduction of computational complexity at the cost of some performance degradation. In this simplified case, summation operations reduce to the $\max$ and $\max_{\mathrm{Re}}$ operations.

### C. *Gaussian Approximation*

To start with, we examine the Gaussian pdf, i.e.,

$$f(x) = \exp\left(-|x|^2\right). \tag{42}$$

Although this does not well approximate ordinary signal constellations such as QAM and PSK, with this the integrals are guaranteed to become a simple closed-form as demonstrated in deriving the linear MMSE from the conditional mean estimation in Section III-C.

The substitution of (42) into (26) and (27) yields

$$\begin{aligned}\alpha(x_2) &= \int_{\mathbb{C}} x\lambda(x,x_2)\exp(-|x|^2)dx \\ &= C\exp\left(\frac{2\mathrm{Re}[wx_2]-v|x_2|^2+|z|^2/(u+N_0)}{N_0}\right)\frac{z^*}{u+N_0}. \qquad (43)\\ \beta(x_2) &= \int_{\mathbb{C}} \lambda(x,x_2)\exp(-|x|^2)dx \\ &= C\exp\left(\frac{2\mathrm{Re}[wx_2]-v|x_2|^2+|z|^2/(u+N_0)}{N_0}\right), \qquad (44)\end{aligned}$$

*1) Type-I:* Substituting (43) and (44) into (25), we obtain the Type-I detection formula as

$$\hat{X}_1 = (u+N_0)^{-1}\frac{\sum_{x_2\in\mathcal{X}_M}\exp\left(\frac{2\mathrm{Re}[wx_2]-v|x_2|^2+|z|^2/(u+N_0)}{N_0}\right)\cdot z^*}{\sum_{x_2\in\mathcal{X}_M}\exp\left(\frac{2\mathrm{Re}[wx_2]-v|x_2|^2+|z|^2/(u+N_0)}{N_0}\right)}. \qquad (45)$$

In the log domain, we obtain

$$\begin{aligned}\log\hat{X}_1 = &-\log(u+N_0) + \sum_{x_2\in\mathcal{X}_M}^{\mathbb{C},\log}\left(\frac{2\mathrm{Re}[wx_2]-v|x_2|^2+|z|^2/(u+N0)}{N_0}+\log r_z - j\phi_z\right) \\ &- \sum_{x_2\in\mathcal{X}_M}^{\mathbb{R},\log}\frac{2\mathrm{Re}[wx_2]-v|x_2|^2+|z|^2/(u+N_0)}{N_0}. \qquad (46)\end{aligned}$$

*2) Type-II:* Similarly, from (28), we obtain the Type-II detection formula as

$$\hat{X}_1 = \frac{\sum_{x_1\in\mathcal{X}_M}\exp\left(\frac{2\mathrm{Re}[wx_2]-v|x_2|^2+|z|^2/(u+N_0)}{N_0}\right)\cdot x_1}{\sum_{x_1\in\mathcal{X}_M}\exp\left(\frac{2\mathrm{Re}[wx_2]-v|x_2|^2+|z|^2/(u+N_0)}{N_0}\right)} \qquad (47)$$

Its log domain form is given by

$$\begin{aligned}\log\hat{X}_1 = &\sum_{x_1\in\mathcal{X}_M}^{\mathbb{C},\log}\left(\frac{2\mathrm{Re}[wx_1]-v|x_1|^2+|z|^2/(u+N_0)}{N_0}+\log|x_1| + j\arg x_1\right) \\ &- \sum_{x_1\in\mathcal{X}_M}^{\mathbb{R},\log}\frac{2\mathrm{Re}[wx_1]-v|x_1|^2+|z|^2/(u+N_0)}{N_0}. \qquad (48)\end{aligned}$$

### *D. Uniform Square Approximation for QAM*

In this subsection, we consider a uniform square approximation which is suitable for rectangular $M$-QAM, where $M$ is an even power of 2. In this approximation, infinite number of symbol points are uniformly distributed in the square region $D_\kappa$ with the vertices at $\kappa + j\kappa$, $-\kappa + j\kappa$, $-\kappa - j\kappa$, $\kappa - j\kappa$. By taking the limit of $M \to \infty$ in the constellation of $M$-QAM:

$$\mathcal{X}_M = \left\{ \frac{-(\sqrt{M}-1)+2n}{\sqrt{2(M-1)/3}} + j\frac{-(\sqrt{M}-1)+2k}{\sqrt{2(M-1)/3}},\ 0 \le n,k < \sqrt{M} \right\}, \tag{49}$$

we find that $\kappa = \sqrt{3/2}$. Using the unit step function, denoted by $U(\cdot)$, $f(x)$ is expressed as

$$f(x) = \frac{1}{(2\kappa)^2}\left\{U(\mathrm{Re}[x]+\kappa) - U(\mathrm{Re}[x]-\kappa)\right\}\left\{U(\mathrm{Im}[x]+\kappa) - U(\mathrm{Im}[x]-\kappa)\right\}. \tag{50}$$

The substitution of (50) into (26) and (27) yields

$$\begin{aligned}\alpha(x_2) &= \frac{1}{(2\kappa)^2}\int_{D_\kappa} x\lambda(x,x_2)dx \\ &= C\exp\left(\frac{2\mathrm{Re}[wx_2] - v|x_2|^2 + |z|^2/u}{N_0}\right) \\ &\cdot \left\{ \left( \sqrt{\frac{N_0}{u\pi}}\left(e^{-\tau_{\mathrm{I}}^{(+)2}} - e^{-\tau_{\mathrm{I}}^{(-)2}}\right) + \mathrm{Re}\left[\frac{z^*}{u}\right]e_{\mathrm{I}} \right) e_{\mathrm{Q}} + j\left( \sqrt{\frac{N_0}{u\pi}}\left(e^{-\tau_{\mathrm{Q}}^{(+)2}} - e^{-\tau_{\mathrm{Q}}^{(-)2}}\right) + \mathrm{Im}\left[\frac{z^*}{u}\right]e_{\mathrm{Q}} \right) e_{\mathrm{I}} \right\},\end{aligned} \tag{51}$$

$$\begin{aligned}\beta(x_2) &= \frac{1}{(2\kappa)^2}\int_{D_\kappa} \lambda(x,x_2)dx \\ &= C\exp\left(\frac{2\mathrm{Re}[wx_2] - v|x_2|^2 + |z|^2/u}{N_0}\right)e_{\mathrm{I}}e_{\mathrm{Q}},\end{aligned} \tag{52}$$

where

$$\tau_{\mathrm{I}}^{(\pm)} := \sqrt{\frac{u}{N_0}}\left(\mathrm{Re}\left[\frac{z^*}{u}\right] \pm \kappa\right), \quad \tau_{\mathrm{Q}}^{(\pm)} := \sqrt{\frac{u}{N_0}}\left(\mathrm{Im}\left[\frac{z^*}{u}\right] \pm \kappa\right), \tag{53}$$

$$e_{\mathrm{I}} := \mathrm{erf}\left(\tau_{\mathrm{I}}^{(+)}\right) - \mathrm{erf}\left(\tau_{\mathrm{I}}^{(-)}\right), \quad e_{\mathrm{Q}} := \mathrm{erf}\left(\tau_{\mathrm{Q}}^{(+)}\right) - \mathrm{erf}\left(\tau_{\mathrm{Q}}^{(-)}\right). \tag{54}$$

*1) Type-I:* In (51), $\sqrt{\frac{N_0}{u\pi}}\left(e^{-\tau_{\mathrm{I}}^{(+)2}} - e^{-\tau_{\mathrm{I}}^{(-)2}}\right)$, $\sqrt{\frac{N_0}{u\pi}}\left(e^{-\tau_{\mathrm{Q}}^{(+)2}} - e^{-\tau_{\mathrm{Q}}^{(-)2}}\right)$ are found to be negligible even when signal-to-noise ratio (SNR) is small. Substituting (51) and (52) into (25) with dropping the negligible terms, we obtain

$$\hat{X}_1 = u^{-1}\frac{\sum_{x_2\in\mathcal{X}_M}\exp\left(\frac{2\mathrm{Re}[wx_2]-v|x_2|^2+|z|^2/u}{N_0}\right)e_{\mathrm{I}}e_{\mathrm{Q}}\cdot z^*}{\sum_{x_2\in\mathcal{X}_M}\exp\left(\frac{2\mathrm{Re}[wx_2]-v|x_2|^2+|z|^2/u}{N_0}\right)e_{\mathrm{I}}e_{\mathrm{Q}}}. \tag{55}$$

Since $e_{\mathrm{I}}$ and $e_{\mathrm{Q}}$ are positive, it can be expressed in the log domain as follows:

$$\begin{aligned}\log(\hat{X}_1) = -\log u &+ \sum_{x_2\in\mathcal{X}_M}^{\mathbb{C},\log}\left(\frac{2\mathrm{Re}[wx_2] - v|x_2|^2 + |z|^2/u}{N_0} + \log e_{\mathrm{I}} + \log e_{\mathrm{Q}} + \log r_z - j\phi_z\right) \\ &- \sum_{x_2\in\mathcal{X}_M}^{\mathbb{R},\log}\left(\frac{2\mathrm{Re}[wx_2] - v|x_2|^2 + |z|^2/u}{N_0} + \log e_{\mathrm{I}} + \log e_{\mathrm{Q}}\right).\end{aligned} \tag{56}$$

*2) Type-II:* Substituting (52) into (28), we obtain the Type-II detection formula as

$$\hat{X}_1 = \frac{\sum_{x_1\in\mathcal{X}_M} \exp\left(\frac{2\mathrm{Re}[wx_2]-v|x_2|^2+|z|^2/u}{N_0}\right) e_{\mathrm{I}} e_{\mathrm{Q}} \cdot x_1}{\sum_{x_1\in\mathcal{X}_M} \exp\left(\frac{2\mathrm{Re}[wx_2]-v|x_2|^2+|z|^2/u}{N_0}\right) e_{\mathrm{I}} e_{\mathrm{Q}}}. \tag{57}$$

Its log domain form is given by

$$\log \hat{X}_1 = \sum_{x_1\in\mathcal{X}_M}^{\mathbb{C},\log} \left(\frac{2\mathrm{Re}[wx_1]-v|x_1|^2+|z|^2/u}{N_0} + \log e_{\mathrm{I}} + \log e_{\mathrm{Q}} + \log|x_1| + j\arg x_1\right)$$
$$- \sum_{x_1\in\mathcal{X}_M}^{\mathbb{R},\log} \left(\frac{2\mathrm{Re}[wx_1]-v|x_1|^2+|z|^2/u}{N_0} + \log e_{\mathrm{I}} + \log e_{\mathrm{Q}}\right). \tag{58}$$

### *E. Uniform Ring Approximation for PSK/APSK*

In this subsection, we consider a uniform ring approximation suitable for $M$-APSK whose constellation is in general expressed in terms of the number of rings $K$, the radii of the rings $\rho_1, \ldots, \rho_K$, and the number of symbol points on each ring $M_k$, $k = 1, \ldots, K$, where $\sum_{k=1}^{K} M_k = M$. With these parameters, the constellation is defined as

$$\mathcal{X}_M = \left\{\rho_k e^{j2\pi n/M_k},\, 0 \le n < M_k, k = 1, \ldots, K\right\}. \tag{59}$$

Note that $M$-PSK corresponds to the case with $K = 1$, $\rho_1 = 1$, $M_1 = M$. In the uniform ring approximation, infinite number of symbol points are uniformly distributed on each ring. Thus,

$$f(x) = \frac{1}{2\pi K}\sum_{k=1}^{K} \delta(|x| - \rho_k), \tag{60}$$

where $\delta(\cdot)$ is the Dirac's delta function.

With $f(x)$ given as (60), it is easier to develop (26) and (27) in polar coordinates by replacing $x$ with $x = \rho_k e^{j\theta}$. It follows that

$$\alpha(x_2) := \frac{1}{2\pi K}\sum_{k=1}^{K} \rho_k \int_{-\pi}^{\pi} e^{j\theta}\lambda(\rho_k e^{j\theta}, x_2) d\theta$$
$$= C\sum_{k=1}^{K} \exp\left(\frac{2\mathrm{Re}[wx_2] - u\rho_k^2 - v|x_2|^2}{N_0}\right) \rho_k e^{-j\phi_z} I_1\left(\frac{2\rho_k r_z}{N_0}\right), \tag{61}$$

$$\beta(x_2) := \frac{1}{2\pi K}\sum_{k=1}^{K} \int_{-\pi}^{\pi} \lambda(\rho_k e^{j\theta}, x_2) d\theta$$
$$= C\sum_{k=1}^{K} \exp\left(\frac{2\mathrm{Re}[wx_2] - u\rho_k^2 - v|x_2|^2}{N_0}\right) I_0\left(\frac{2\rho_k r_z}{N_0}\right), \tag{62}$$

where $I_0(\cdot)$ and $I_1(\cdot)$ are respectively the zeroth and first order modified Bessel functions of the first kind. These Bessel functions can be well approximated as $I_0(x) \approx \exp(x)$ and $I_1(x) \approx \exp(x)$.

*1) Type-I:* Substituting (61) and (62) into (28) with each Bessel function replaced by its exponential approximation form, we obtain

$$\hat{X}_1 = \frac{\sum_{x_2 \in \mathcal{X}_M} \sum_{k=1}^{K} \rho_k e^{-j\phi_z} \exp\left(\frac{2(\mathrm{Re}[wx_2]+\rho_k r_z)-u\rho_k^2-v|x_2|^2}{N_0}\right)}{\sum_{x_2 \in \mathcal{X}_M} \sum_{k=1}^{K} \exp\left(\frac{2(\mathrm{Re}[wx_2]+\rho_k r_z)-u\rho_k^2-v|x_2|^2}{N_0}\right)}, \tag{63}$$

whose log domain form is

$$\log \hat{X}_1 = \sum_{x_2 \in \mathcal{X}_M}^{\mathbb{C},\log} \sum_{k=1,\ldots,K}^{\mathbb{C},\log} \left(\frac{2(\mathrm{Re}[wx_2]+\rho_k r_z)-u\rho_k^2-v|x_2|^2+\log\rho_k}{N_0} - j\phi_z\right)$$
$$- \sum_{x_2 \in \mathcal{X}_M}^{\mathbb{R},\log} \sum_{k=1,\ldots,K}^{\mathbb{R},\log} \frac{2(\mathrm{Re}[wx_2]+\rho_k r_z)-u\rho_k^2-v|x_2|^2}{N_0}. \tag{64}$$

As mentioned above, a PSK constellation corresponds to the one with $K=1$ and $\rho_1 = 1$, which simplifies the detection formula as

$$\log \hat{X}_1 = \sum_{x_2 \in \mathcal{X}_M}^{\mathbb{C},\log} \left(\frac{2(\mathrm{Re}[wx_2]+r_z)}{N_0} - j\phi_z\right) - \sum_{x_2 \in \mathcal{X}_M}^{\mathbb{R},\log} \frac{2(\mathrm{Re}[wx_2]+r_z)}{N_0}. \tag{65}$$

*2) Type-II:* The substitution of (62) into (28) yields the Type-II formula as

$$\hat{X}_1 = \frac{\sum_{x_1 \in \mathcal{X}_M} x_1 \sum_{k=1}^{K} \exp\left(\frac{2(\mathrm{Re}[wx_1]+\rho_k r_z)-u\rho_k^2-v|x_1|^2}{N_0}\right)}{\sum_{x_1 \in \mathcal{X}_M} \sum_{k=1}^{K} \exp\left(\frac{2(\mathrm{Re}[wx_1]+\rho_k r_z)-u\rho_k^2-v|x_1|^2}{N_0}\right)}. \tag{66}$$

In the log domain,

$$\log \hat{X}_1 = \sum_{x_1 \in \mathcal{X}_M}^{\mathbb{C},\log} \sum_{k=1,\ldots,K}^{\mathbb{C},\log} \left(\frac{2(\mathrm{Re}[wx_1]+\rho_k r_z)-u\rho_k^2-v|x_1|^2+\log|x_1|}{N_0} + j\arg x_1\right)$$
$$- \sum_{x_1 \in \mathcal{X}_M}^{\mathbb{R},\log} \sum_{k=1,\ldots,K}^{\mathbb{R},\log} \frac{2(\mathrm{Re}[wx_1]+\rho_k r_z)-u\rho_k^2-v|x_1|^2}{N_0}. \tag{67}$$

For the PSK constellations, we obtain

$$\log \hat{X}_1 = \sum_{x_1 \in \mathcal{X}_M}^{\mathbb{C},\log} \left(\frac{2(\mathrm{Re}[wx_1]+r_z)}{N_0} + j\arg x_1\right) - \sum_{x_1 \in \mathcal{X}_M}^{\mathbb{R},\log} \frac{2(\mathrm{Re}[wx_1]+r_z)}{N_0}. \tag{68}$$

## V. Simulation Results

In this section, we evaluate the bit error rate (BER) performance of the proposed signal detection method via computer simulation. The fading coefficients $h_{k,n}$, $k = 1, \ldots, N_r$, $n = 1, 2$ are assumed to be mutually-independent complex Gaussian RVs each with zero mean and unit variance. The SNR per information bit is denoted by $E_b/N_0$.

Figures 2, 3, and 4 contain the plot of BER curves of QPSK, 8-PSK, and 16-PSK detected based on the uniform ring approximation, where it can be seen that the performances of both Type-I and II gradually deviate from that of the linear MMSE, and the Type-I detection always outperforms the Type-II detection. At $\mathrm{BER} = 10^{-4}$, the required SNR is improved by 11 or 12 dB over the linear MMSE in the case of Type-I, even approaching the performance of MLD within 3 dB. In these figures, the performances of the Type-I and II detections, where the Jacobian logarithm summations are replaced by the $\max$ and $\max_{\mathrm{Re}}$ operations (see Section IV-B), are also plotted (labeled with "max" in the legend). The performance gap between these approaches is as small as 0.5 dB in the cases of QPSK and 8-PSK with Type-I detection, and even becomes negligible in the other cases. We thus remark that this simplification of the Jacobian summations is a preferable option in practice, as it significantly reduces complexity.

In Fig. 5, we make a comparison of BER performances with increasing $N_r$ from 2 to 4, for 8-PSK detected by the Type-I uniform ring approximation. As observed, increasing $N_r$ remarkably reduces the SNR gaps from MLD: there is a loss of as small as 0.8 dB in $N_r = 3$, and there is no noticeable loss in $N_r = 4$.

In Fig. 6, we plot the BER performance of a 16-APSK constellation with the parameters of $K = 2$, $\rho_1 = \sqrt{2/5}$, $\rho_2 = 2\sqrt{2/5}$, and $M_1 = M_2 = 8$, which corresponds to a popular setting of the 16-APSK consisting of two concentric 8-PSK with the ring ratio of $1:2$. We can observe in the BER curves a tendency similar to that of the PSKs. The performance gap between Type-I and MLD is, however, slightly larger than that in the PSK cases.

The detection based on the uniform square approximation is evaluated in Fig. 7 with $N_r = 2$ and 16-QAM as an example. Similar to the PSK/APSK cases, Type-I clearly outperforms Type-II, but their BER curves show that achievable diversity order is not as large as that in PSK/APSK. As a result, compared with the linear MMSE, the SNR improvement at $\mathrm{BER} = 10^{-4}$ is limited to 3.5 dB even in Type-I.

Unlike the uniform square and ring approximations, the Gaussian one does not target a specific constellation. We therefore examine its applicability to both 8-PSK and 16-QAM. Figure 8 compares the Type-I detection of 8-PSK with both Gaussian and uniform ring approximation. Noticeable gap is observed, suggesting the considerable dependence of the symbol distribution approximation on the overall performance. Shown in Fig. 9 is the case for 16-QAM, Type-I, with both Gaussian and uniform square approximation. In the latter case, the gap becomes small, but the square approximation still outperforms Gaussian.

Lastly, as a simple demonstration of low-complexity aspect of the proposed detection framework, in Table II we compare average processing time for detecting 1,000,000 symbols in our simulation program written in C++. Here, we focus only on the PSKs with the uniform ring approximation which has a BER performance close to that of MLD. As a reference, in the table the number of multiplications[1] for iteration and the number of iterations required for detecting one symbol are also listed. As observed, the processing time of the Type-I and II detections are significantly lower than that of MLD.

## VI. Conclusion

In this paper, we have proposed a low-complexity detection scheme for MIMO-SM systems, based on the generalized MMSE detection derived from the conditional mean estimation expression. Our main focus is on the case with two transmit antennas, but the proposed scheme can be extended to the general system having any number of transmit antennas in a straightforward manner. The reduction of complexity is due to the approximation of the probability distribution in the conditional mean estimation expression such that the resulting formula in part has a simple closed form. The proposed detection has two sub-classes: in Type-I the desired symbol is approximated, whereas the approximation is in turn applied to the interfering symbol in Type-II. By simulation, we have observed that the Type-I detection is superior to Type-II. Furthermore, in the case of PSK and APSK, the former detection with the uniform ring approximation achieves a BER performance even comparable to MLD. Finally, as a possible future work, we note that there may be a further room of complexity reduction and BER improvement by utilizing approximations other than those examined in this paper.

[1]Note that the number of *real* multiplications is displayed. A complex multiplication accounts for 4 real multiplications, but it is reduced to 2 when followed by $\mathrm{Re}$. An absolute square of a complex number takes 2 real multiplications. For MLD, the multiplication count is half the actual one since it can detect $X_1$ and $X_2$ simultaneously. For the Type-I detection, we note that computational burden is also on decomposing a complex number into its phase and amplitude, whereas for Type-II, the phase is not necessary and thus its processing time is lower than that of Type-I.

## Appendix I
## Equivalence of Two Expressions of Linear MMSE

A proof of the equivalence between (10) and (11) proceeds as follows:

$$\begin{aligned}
&\left(\boldsymbol{H}^H\boldsymbol{H}+N_0\boldsymbol{I}_{N_t}\right)^{-1}\boldsymbol{H}^H\\
=&\left\{\frac{1}{N_0}\boldsymbol{I}_{N_t}-\frac{1}{N_0}\boldsymbol{H}^H\left(\frac{1}{N_0}\boldsymbol{H}\boldsymbol{H}^H+\boldsymbol{I}_{N_r}\right)^{-1}\frac{1}{N_0}\boldsymbol{H}\right\}\boldsymbol{H}^H\\
=&\frac{1}{N_0}\boldsymbol{H}^H-\frac{1}{N_0}\boldsymbol{H}^H\left(\frac{1}{N_0}\boldsymbol{H}\boldsymbol{H}^H+\boldsymbol{I}_{N_r}\right)^{-1}\boldsymbol{H}\boldsymbol{H}^H\\
=&\frac{1}{N_0}\boldsymbol{H}^H\left(\frac{1}{N_0}\boldsymbol{H}\boldsymbol{H}^H+\boldsymbol{I}_{N_r}\right)^{-1}\left(\frac{1}{N_0}\boldsymbol{H}\boldsymbol{H}^H+\boldsymbol{I}_{N_r}\right)-\frac{1}{N_0}\boldsymbol{H}^H\left(\frac{1}{N_0}\boldsymbol{H}\boldsymbol{H}^H+\boldsymbol{I}_{N_r}\right)^{-1}\frac{1}{N_0}\boldsymbol{H}\boldsymbol{H}^H\\
=&\frac{1}{N_0}\boldsymbol{H}^H\left(\frac{1}{N_0}\boldsymbol{H}\boldsymbol{H}^H+\boldsymbol{I}_{N_r}\right)^{-1}\left(\frac{1}{N_0}\boldsymbol{H}\boldsymbol{H}^H+\boldsymbol{I}_{N_r}-\frac{1}{N_0}\boldsymbol{H}\boldsymbol{H}^H\right)\\
=&\boldsymbol{H}^H\left\{N_0\left(\frac{1}{N_0}\boldsymbol{H}\boldsymbol{H}^H+\boldsymbol{I}_{N_r}\right)\right\}^{-1}\boldsymbol{I}_{N_r}\\
=&\boldsymbol{H}^H\left(\boldsymbol{H}\boldsymbol{H}^H+N_0\boldsymbol{I}_{N_r}\right)^{-1}.
\end{aligned}\tag{69}$$

## Appendix II
## Proof of Proposition 1

Substituting $p_{\boldsymbol{X}}(\boldsymbol{x})=\frac{1}{\pi^{N_t}}e^{-||\boldsymbol{x}||^2}$ into (19), we obtain

$$\begin{aligned}
\hat{\boldsymbol{X}}&=\frac{\int_{\mathbb{C}^{N_t}}\boldsymbol{x}\cdot\exp\left(-\frac{||\boldsymbol{y}-\boldsymbol{H}\boldsymbol{x}||^2+N_0||\boldsymbol{x}||^2}{N_0}\right)d^{N_t}\boldsymbol{x}}{\int_{\mathbb{C}^{N_t}}\exp\left(-\frac{||\boldsymbol{y}-\boldsymbol{H}\boldsymbol{x}||^2+N_0||\boldsymbol{x}||^2}{N_0}\right)d^{N_t}\boldsymbol{x}}\\
&=\frac{\int_{\mathbb{C}^{N_t}}\boldsymbol{x}\cdot\exp\left(-\frac{-\boldsymbol{y}^H\boldsymbol{H}\boldsymbol{x}-(\boldsymbol{y}^H\boldsymbol{H}\boldsymbol{x})^*+\boldsymbol{x}^H\boldsymbol{H}^H\boldsymbol{H}\boldsymbol{x}+N_0\boldsymbol{x}^H\boldsymbol{x}}{N_0}\right)d^{N_t}\boldsymbol{x}}{\int_{\mathbb{C}^{N_t}}\exp\left(-\frac{-\boldsymbol{y}^H\boldsymbol{H}\boldsymbol{x}-(\boldsymbol{y}^H\boldsymbol{H}\boldsymbol{x})^*+\boldsymbol{x}^H\boldsymbol{H}^H\boldsymbol{H}\boldsymbol{x}+N_0\boldsymbol{x}^H\boldsymbol{x}}{N_0}\right)d^{N_t}\boldsymbol{x}}.
\end{aligned}\tag{70}$$

Let $\boldsymbol{Q}$ be the unitary matrix that diagonalizes the Hermitian matrix $\boldsymbol{H}^H\boldsymbol{H}$ as

$$\boldsymbol{Q}\boldsymbol{H}^H\boldsymbol{H}\boldsymbol{Q}^H=\boldsymbol{\Lambda},\tag{71}$$

where $\boldsymbol{\Lambda}$ is a diagonal matrix that gathers the eigenvalues of $\boldsymbol{H}^H\boldsymbol{H}$, $\lambda_i$, $1 \leq i \leq N_t$. With the unitary matrix $\boldsymbol{Q}$, we translate the problem into finding $\hat{\boldsymbol{Z}} := \boldsymbol{Q}\hat{\boldsymbol{X}}$ instead of $\hat{\boldsymbol{X}}$. Changing the variable $\boldsymbol{x}$ to $\boldsymbol{Q}^H\boldsymbol{x}$ in (70) then yields $\hat{\boldsymbol{Z}}$ as

$$\hat{\boldsymbol{Z}} = \frac{\int_{\mathbb{C}^{N_t}} \boldsymbol{x} \cdot \exp\left(-\frac{(-\boldsymbol{y}^H\boldsymbol{H}\boldsymbol{Q}^H\boldsymbol{x} - (\boldsymbol{y}^H\boldsymbol{H}\boldsymbol{Q}^H\boldsymbol{x})^* + \boldsymbol{x}^H(\boldsymbol{\Lambda}+N_0\boldsymbol{I})\boldsymbol{x})}{N_0}\right) d^{N_t}\boldsymbol{x}}{\int_{\mathbb{C}^{N_t}} \exp\left(-\frac{(-\boldsymbol{y}^H\boldsymbol{H}\boldsymbol{Q}^H\boldsymbol{x} - (\boldsymbol{y}^H\boldsymbol{H}\boldsymbol{Q}^H\boldsymbol{x})^* + \boldsymbol{x}^H(\boldsymbol{\Lambda}+N_0\boldsymbol{I})\boldsymbol{x})}{N_0}\right) d^{N_t}\boldsymbol{x}}. \tag{72}$$

Let $a_i$ denote the $i$-th element in the row vector $\boldsymbol{y}^H\boldsymbol{H}\boldsymbol{Q}^H$. The $i$-th element in $\hat{\boldsymbol{Z}}$, denoted by $\hat{Z}_i$, is then written as

$$\hat{Z}_i = \frac{\int_{\mathbb{C}} x \cdot \exp\left(-\frac{-a_i x - a_i^* x^* + (\lambda_i + N_0)xx^*}{N_0}\right) dx \prod_{n=1,n\neq i}^{N_t} \int_{\mathbb{C}} \exp\left(-\frac{-a_n x - a_n^* x^* + (\lambda_n + N_0)xx^*}{N_0}\right) dx}{\prod_{n=1}^{N_t} \int_{\mathbb{C}} \exp\left(-\frac{-a_n x - a_n^* x^* + (\lambda_n + N_0)xx^*}{N_0}\right) dx}. \tag{73}$$

Noticing that

$$\begin{aligned} &\int_{\mathbb{C}} \exp\left(-\frac{-a_n x - a_n^* x^* + (\lambda_n + N_0)xx^*}{N_0}\right) dx \\ &= \int_{\mathbb{C}} \exp\left(-\frac{\lambda_n + N_0}{N_0}\left(\left|x - \frac{a_n^*}{(\lambda_n + N_0)}\right|^2 - \frac{|a_n|^2}{(\lambda_n + N_0)^2}\right)\right) dx \\ &= \exp\left(\frac{|a_n|^2}{N_0(\lambda_n + N_0)}\right) \frac{\pi N_0}{\lambda_n + N_0} \end{aligned} \tag{74}$$

and that

$$\begin{aligned} &\int_{\mathbb{C}} x \cdot \exp\left(-\frac{-a_i x - a_i^* x^* + (\lambda_i + N_0)xx^*}{N_0}\right) dx \\ &= \exp\left(\frac{|a_i|^2}{N_0(\lambda_i + N_0)}\right) \frac{\pi N_0}{\lambda_i + N_0} \frac{a_i^*}{\lambda_i + N_0}. \end{aligned} \tag{75}$$

(73) can be simplified as we obtain

$$\hat{Z}_i = \frac{a_i^*}{\lambda_i + N_0}. \tag{76}$$

Thus,

$$\hat{\boldsymbol{Z}} = \begin{bmatrix} \frac{a_1^*}{\lambda_1 + N_0} \\ \vdots \\ \frac{a_{N_t}^*}{\lambda_{N_t} + N_0} \end{bmatrix} = (\boldsymbol{\Lambda} + N_0\boldsymbol{I}_{N_t})^{-1}(\boldsymbol{y}^H\boldsymbol{H}\boldsymbol{Q}^H)^H = (\boldsymbol{\Lambda} + N_0\boldsymbol{I}_{N_t})^{-1}\boldsymbol{Q}\boldsymbol{H}^H\boldsymbol{y}. \tag{77}$$

Finally we invert the transformation $\hat{\boldsymbol{Z}} = \boldsymbol{Q}\hat{\boldsymbol{X}}$ to obtain

$$\begin{aligned}\hat{\boldsymbol{X}} &= \boldsymbol{Q}^H\hat{\boldsymbol{Z}} = \boldsymbol{Q}^H(\boldsymbol{\Lambda} + N_0\boldsymbol{I}_{N_t})^{-1}\boldsymbol{Q}\boldsymbol{H}^H\boldsymbol{y} = \boldsymbol{Q}^H(\boldsymbol{Q}\boldsymbol{H}^H\boldsymbol{H}\boldsymbol{Q}^H + N_0\boldsymbol{I}_{N_t})^{-1}\boldsymbol{Q}\boldsymbol{H}^H\boldsymbol{y} \\ &= \boldsymbol{Q}^H(\boldsymbol{Q}(\boldsymbol{H}^H\boldsymbol{H} + N_0\boldsymbol{I}_{N_t})\boldsymbol{Q}^H)^{-1}\boldsymbol{Q}\boldsymbol{H}^H\boldsymbol{y} = \boldsymbol{Q}^H\boldsymbol{Q}^{-H}(\boldsymbol{H}^H\boldsymbol{H} + N_0\boldsymbol{I}_{N_t})^{-1}\boldsymbol{Q}^{-1}\boldsymbol{Q}\boldsymbol{H}^H\boldsymbol{y} \\ &= \underbrace{(\boldsymbol{H}^H\boldsymbol{H} + N_0\boldsymbol{I}_{N_t})^{-1}\boldsymbol{H}^H}_{=\boldsymbol{G}_{\text{MMSE}}}\boldsymbol{y}, \end{aligned} \tag{78}$$

is precisely the linear MMSE detection form of (6) with (11).

## Appendix III
## Jacobian Logarithm

Let $A$ and $B$ be real-valued and greater than zero, and suppose $a := \log A$ and $b := \log B$. The Jacobian logarithm appears as a core formula in the log domain implementation of the MAP decoding algorithm [18], with which one can write $\log(A+B)$ in terms of $a$ and $b$ as

$$\begin{aligned}\log(A+B) &= \log\left(A(1+B/A)\right) = a + \log(1+B/A) = a + \log(1+e^b/e^a) \\ &= a + \log(1+e^{-(a-b)}),\end{aligned} \tag{79}$$

or as

$$\log(A+B) = b + \log(1+e^{-(b-a)}). \tag{80}$$

Notice that in (79) the exponential $e^{-(a-b)} \le 1$ for the case when $a-b \ge 0$, whereas in (80) $e^{-(b-a)} \le 1$ for $a-b<0$. Hence, (79) and (80) are guaranteed not to overflow within the corresponding ranges of $a-b$. Collecting the equations and conditions in a compacted form, we have

$$\log(A+B) = \max(a,b) + \log\left(1+e^{-|a-b|}\right) \tag{81}$$

which never overflows for any values of $a$ and $b$.

The Jacobian logarithm in the real domain presented above is widely known. To extend it to the complex domain, we observe that the equations (79) and (80) are also valid for $a$ and $b$ being complex numbers. The conditions for satisfying $|e^{-(a-b)}| \le 1$ and $|e^{-(b-a)}| < 1$ are given by $\mathrm{Re}[a] - \mathrm{Re}[b] \ge 0$ and $\mathrm{Re}[a] - \mathrm{Re}[b] < 0$, respectively, from which the complex domain Jacobian logarithm follows as

$$\log(A+B) = \max_{\mathrm{Re}}(a,b) + \log\left(1+e^{-(\max_{\mathrm{Re}}(a,b)-\min_{\mathrm{Re}}(a,b))}\right). \tag{82}$$

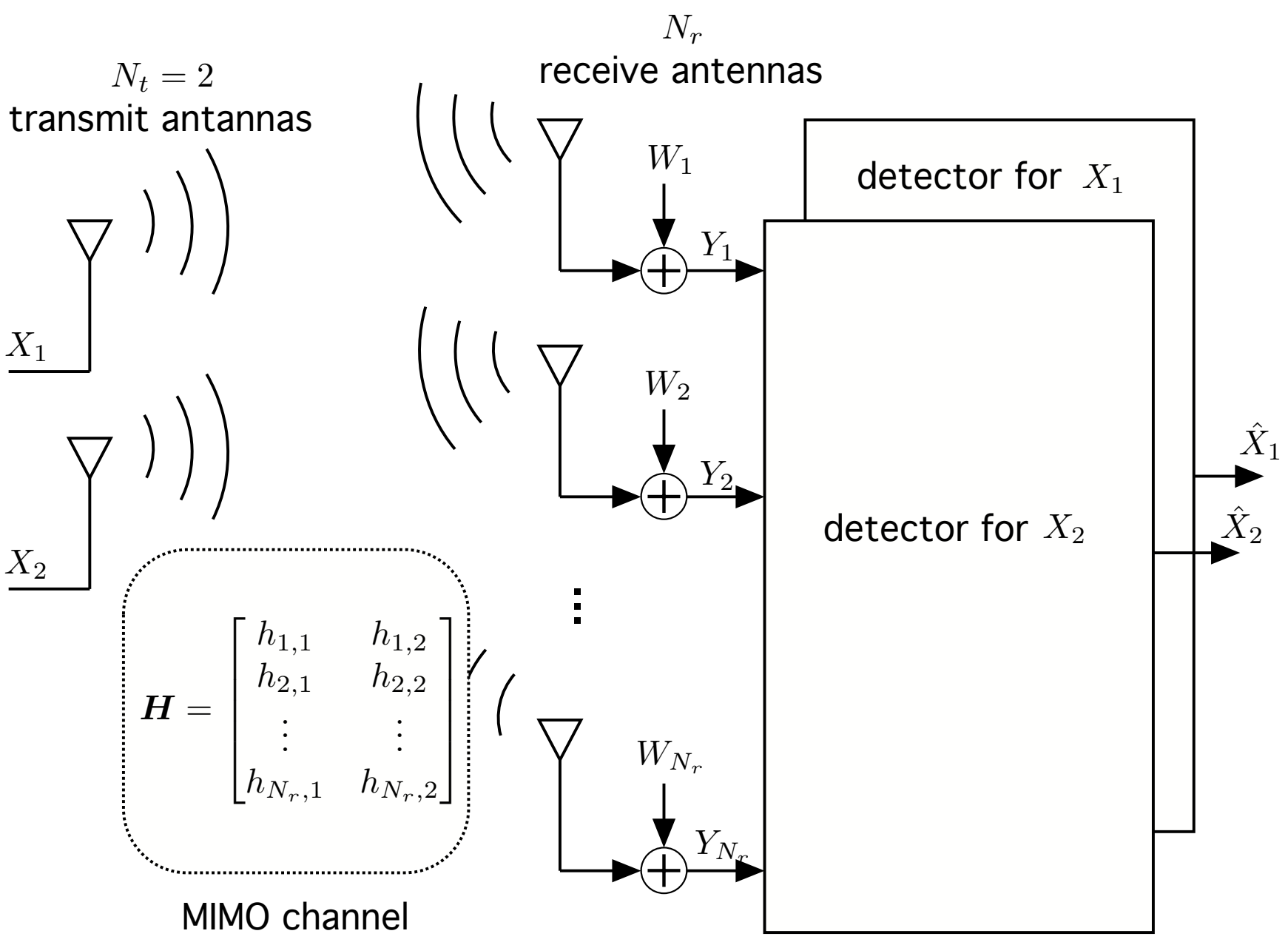

Fig. 1. The proposed MIMO System model. The transmitter with $N_t = 2$ is depicted as an example.

TABLE I

LIST OF THE DERIVED DETECTION FORMULAE.

| Shape of $f(x)$ | Gaussian | 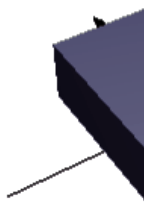 uniform square | 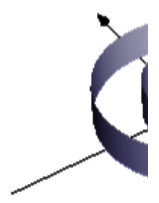 uniform ring | |
|---|---|---|---|---|
| Target constellation | unspecified | QAM | APSK | PSK |
| type-I formula | (46) | (56) | (64) | (65) |
| type-II formula | (48) | (58) | (67) | (68) |

TABLE II

AVERAGE TIME (SEC) FOR PROCESSING 1,000,000 SYMBOLS IN OUR SIMULATION PROGRAM WRITTEN IN C++ ($N_r = 2$).

THE VALUES IN PARENTHESES ARE WITHOUT THE SIMPLIFICATION OF JACOBIAN SUMMATION.

| | Iteration | Multiplication per iter. | QPSK | 8-PSK | 16-PSK |
|---|---|---|---|---|---|
| Linear MMSE | — | $4N_r$ | 0.57 | 0.58 | 0.54 |
| MLD | $M^2$ | $5N_r$ | 1.32 | 2.26 | 5.52 |
| uniform ring type-I | $M$ | $12N_r + 3$ | 0.90 (1.65) | 1.37 (2.95) | 2.27 (5.88) |
| uniform ring type-II | | | 0.71 (1.35) | 1.10 (2.55) | 1.82 (5.03) |

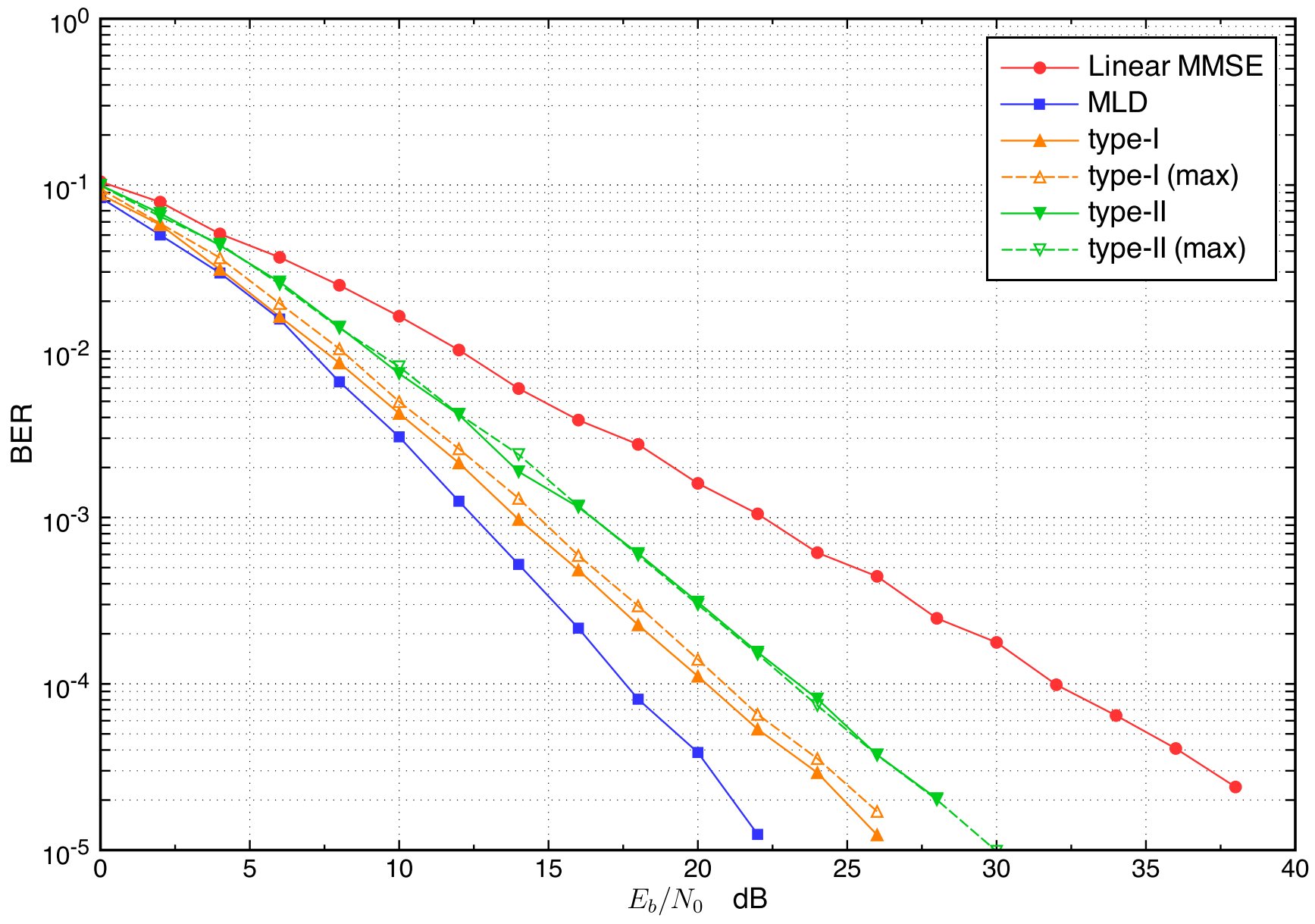

Fig. 2. BER curves of QPSK detected by the Type-I and II formulae with the uniform ring approximation. The number of receive antennas is $N_r = 2$.

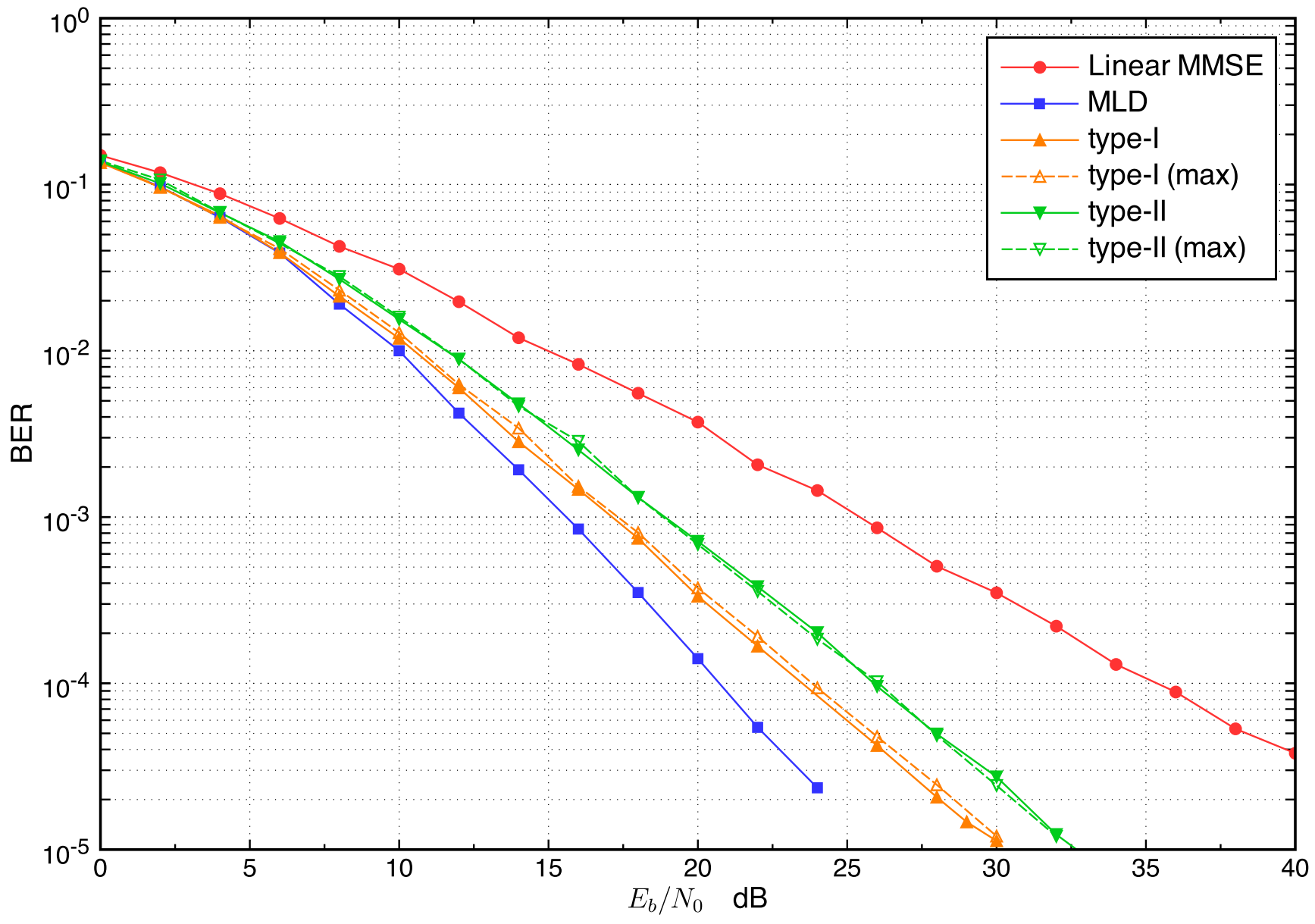

Fig. 3. BER curves of 8-PSK detected by the Type-I and II formulae with the uniform ring approximation. $N_r = 2$.

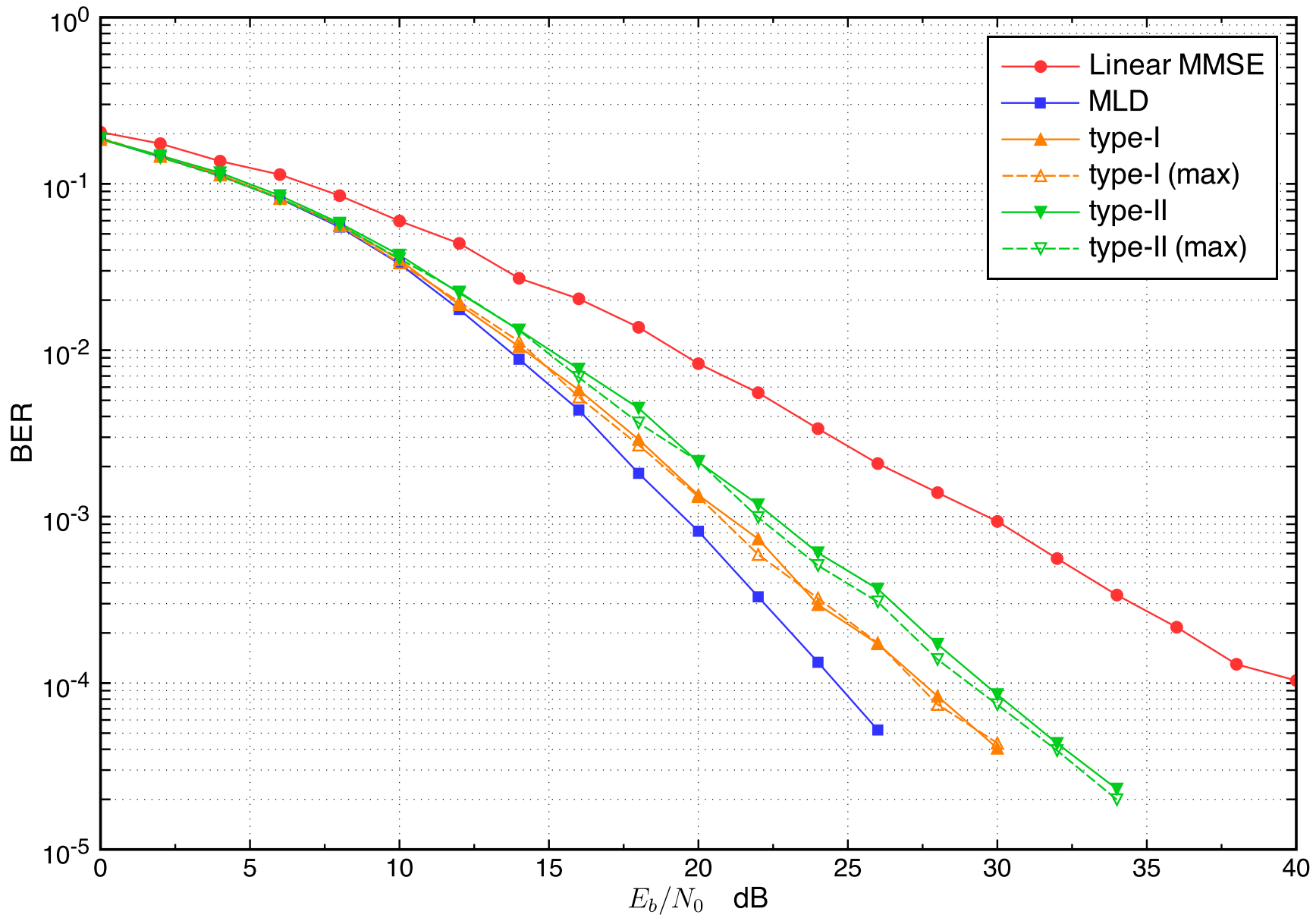

Fig. 4. BER curves of 16-PSK detected by the Type-I and II formulae with the uniform ring approximation. $N_r = 2$.

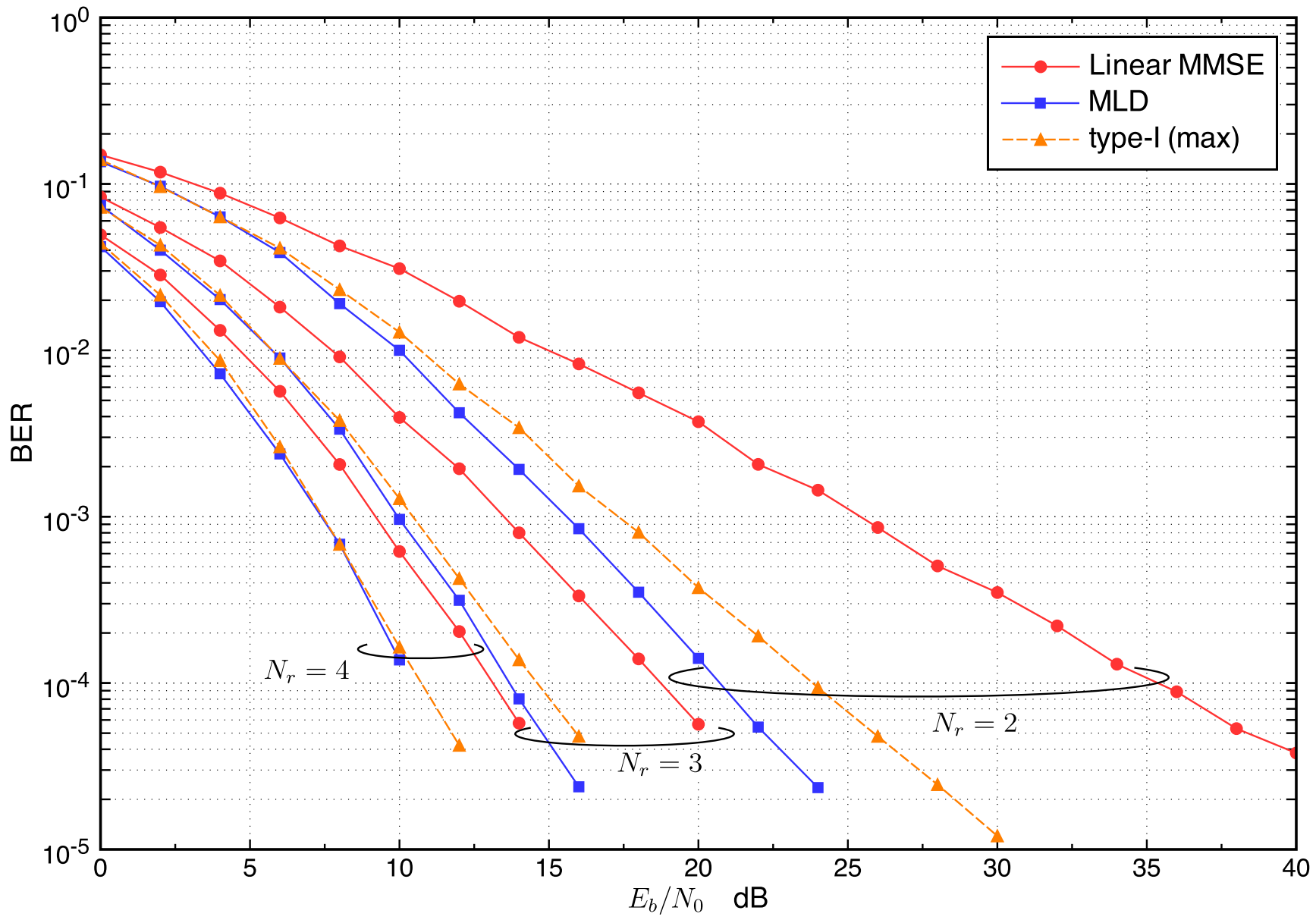

Fig. 5. BER comparison across $N_r = 2, 3,$ and $4$ for 8-PSK with the Type-I uniform ring approximation.

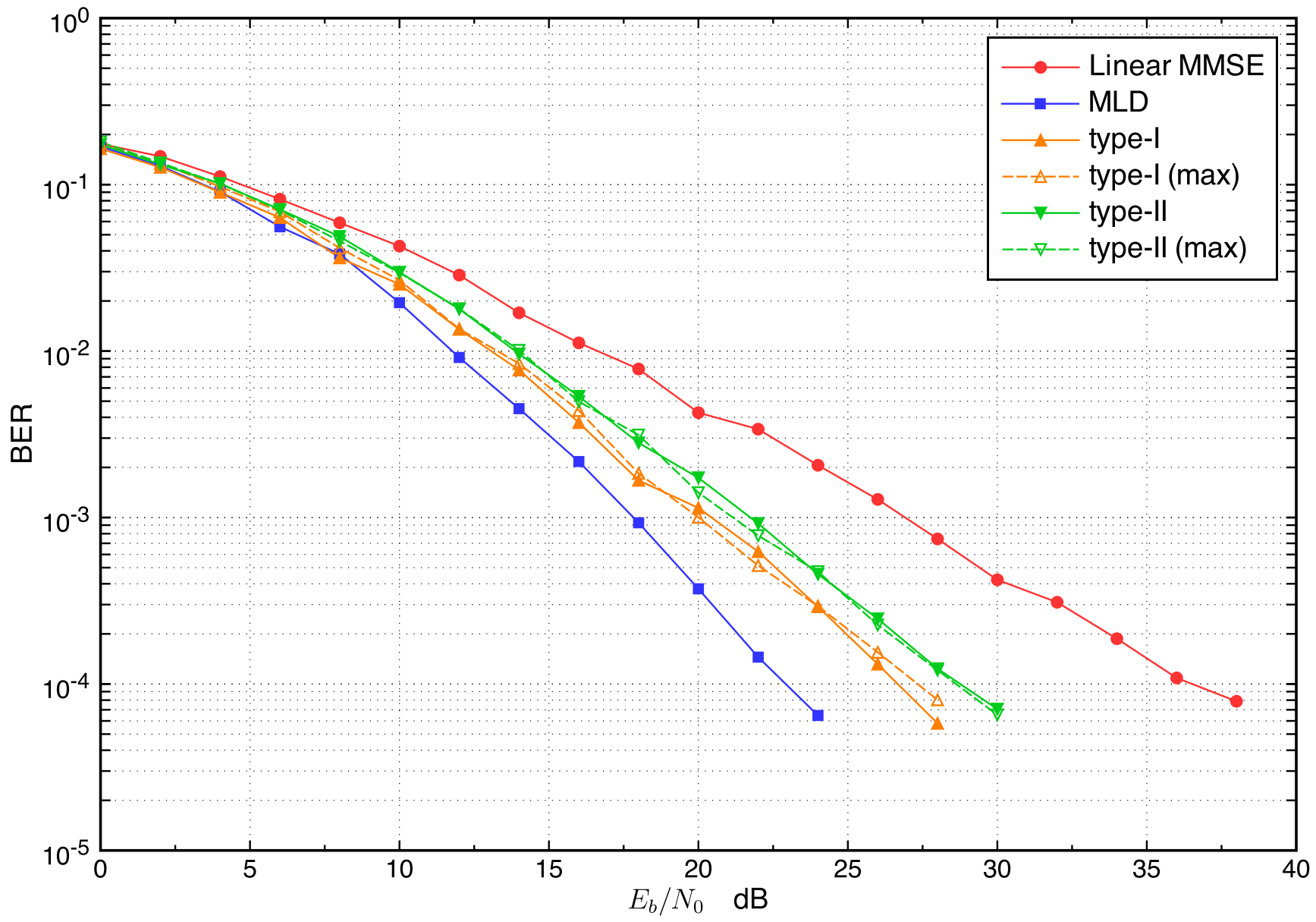

Fig. 6. BER curves of 16-PSK detected by the Type-I and II formulae with the uniform ring approximation. $N_r = 2$.

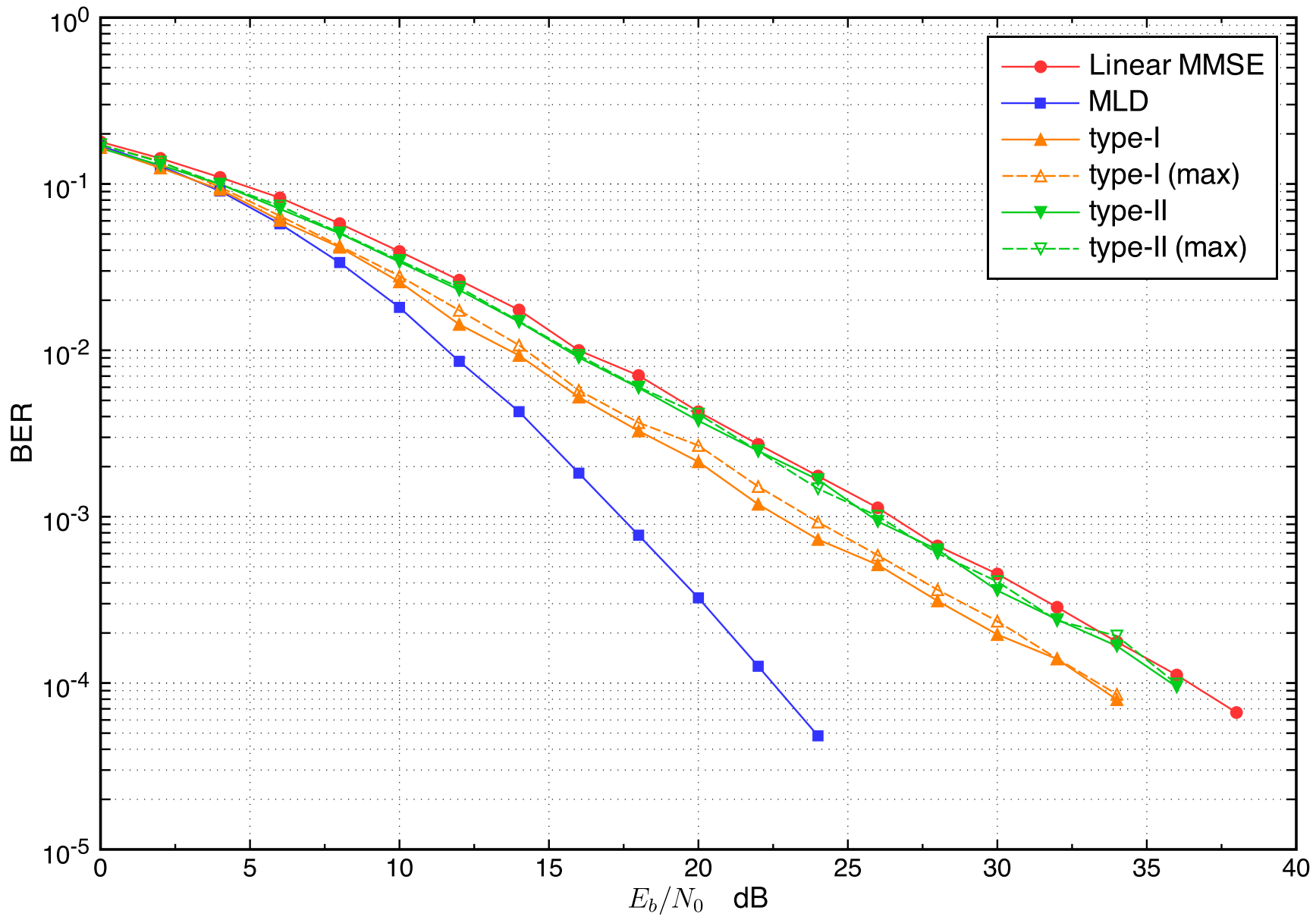

Fig. 7. BER curves of 16-QAM detected by the Type-I and II formulae with the uniform square approximation. $N_r = 2$.

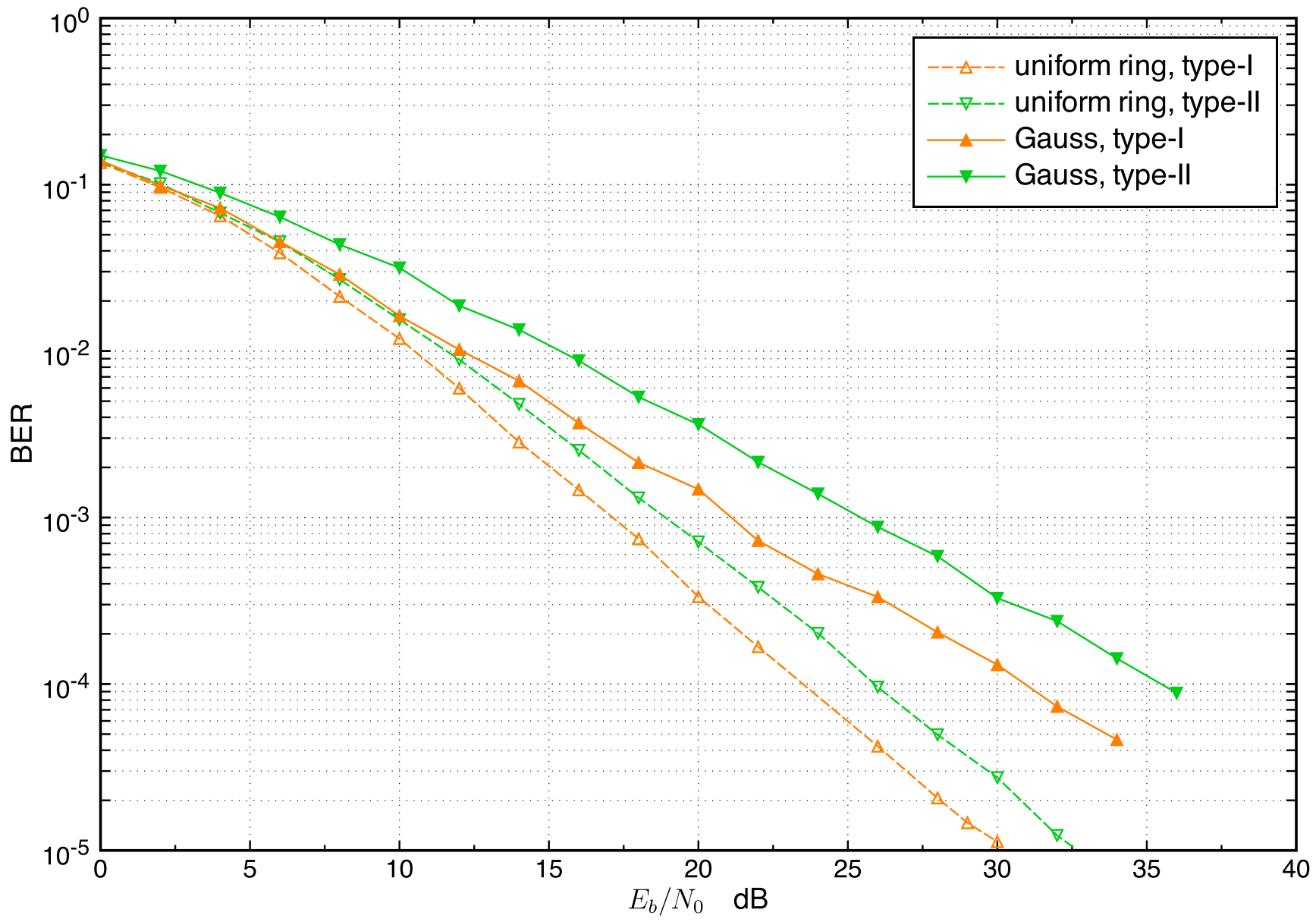

Fig. 8. BER comparison between the Gaussian and uniform ring approximations for 8-PSK. $N_r = 2$.

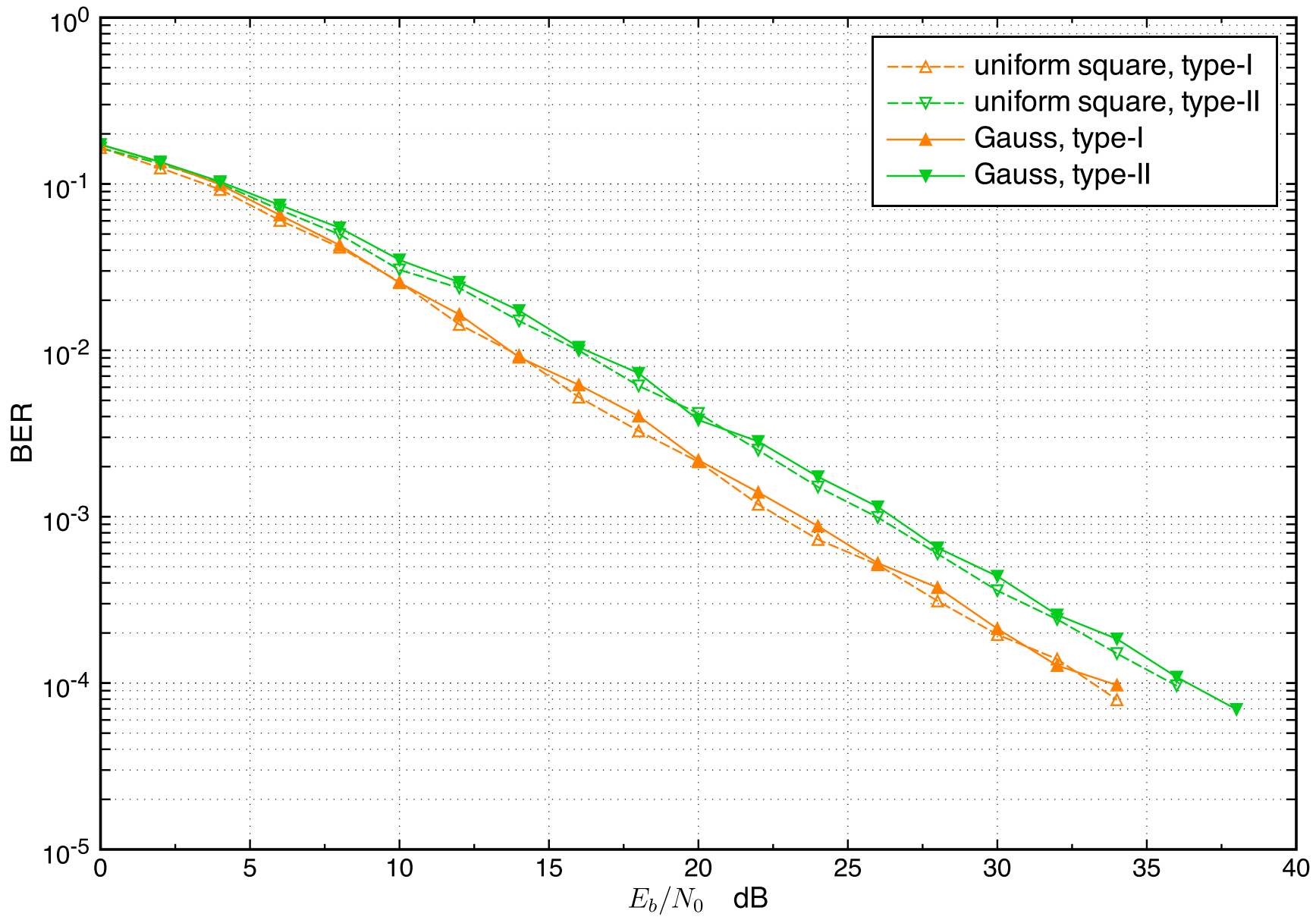

Fig. 9. BER comparison between the Gaussian and uniform square approximations for 16-QAM. $N_r = 2$.